\documentclass[sigconf,nonacm,pdfa]{acmart}

\AtBeginDocument{%
  }

\setcopyright{acmlicensed}
\copyrightyear{2018}
\acmYear{2018}
\acmDOI{XXXXXXX.XXXXXXX}
\acmConference[Conference acronym 'XX]{Make sure to enter the correct
  conference title from your rights confirmation email}{June 03--05,
  2018}{Woodstock, NY}
\acmISBN{978-1-4503-XXXX-X/2018/06}

\setcopyright{none}
\acmConference[]{}{}{}
\copyrightyear{}
\acmYear{}
\acmDOI{}
\acmISBN{}

\usepackage{enumitem}
\usepackage[colorinlistoftodos]{todonotes}
\usepackage{amsmath}
\usepackage{algorithmic}
\usepackage{graphicx}
\usepackage{textcomp}
\usepackage{xcolor}
\usepackage{booktabs}
\usepackage[normalem]{ulem}
\newtheorem{lemma}{Lemma}
\usepackage{tabularx}
\usepackage{array}
\usepackage{comment}
\usepackage{listings}

\begin{document}

\title{Logical Design Tuning via Workload-aware E/R-to-Relational Mapping}

\author{Hasara Kalumin}
\affiliation{%
  \institution{University of Maryland}
  \city{College Park}
  \country{USA}}
\email{hdesilva@umd.edu}

\author{Amol Deshpande}
\affiliation{%
  \institution{University of Maryland}
  \city{College Park}
  \country{USA}
}
\email{amol@umd.edu}


\begin{abstract}
Conceptual data models, like the E/R model, can provide a more stable and intuitive user-facing interface and better logical independence for applications, compared to the relational model which is forced to closely mirror the physical representation. 
However, this flexibility introduces a challenge as well as an opportunity: a single conceptual schema admits many semantically equivalent relational schemas, whose performance can differ substantially across workloads.
Traditional mapping methods use a set of predefined transformation rules or designer-selected mapping strategies, rather than optimizing the relational realization with respect to the expected workload. 
In this paper, we present a cost-based framework for workload-aware conceptual-to-relational mapping. Given an E/R schema and a representative workload, our framework automatically selects an efficient relational schema by optimizing mapping decisions, including inheritance, entity and relationship representations. 
User-specified E/R queries are compiled into SQL over the selected relational realization, allowing existing relational query optimizers to determine the physical execution plan. Experiments on real-world-inspired schemas show that mapping choices strongly affect performance; fixed or heuristic strategies exhibit large variability across workloads, whereas our approach consistently achieves competitive performance avoiding pathological cases.
\end{abstract}



\maketitle

\section{Introduction}

There has been a renewed interest in abstract data models, including Entity-Relationship (E/R) models, property graph models, and object-relational frameworks (ORMs). These higher-level models are closer to user intent, easier to reason about, and have been argued to provide significant benefits in terms of clarity, maintainability, and expressiveness~\cite{deshpande2025beyond,pg-schemas,pg-keys,E-R-graphs}. In many modern systems, the user-facing abstraction is no longer purely relational, but instead reflects a richer conceptual model.
Despite this shift in abstraction, relational database management systems (RDBMSs) remain the dominant storage backends. Consequently, higher-level schemas are typically translated into relational schemas using rule-based mappings. For E/R models, a standard set of textbook rules is used to derive relations, keys, and foreign key constraints~\cite{silberschatz2019database}. Similar rule-based translations have been proposed for XML~\cite{shanmugasundaram1999relational}, property graphs~\cite{opencypher2017, angles2018foundations}, and ORMs~\cite{ambler2003mapping, hibernate}.

The relational model provides solid
physical data independence, but it tightly couples conceptual
abstractions to logical representations, foregoing many performance enhancement opportunities. 
However, once the higher-level abstraction is fixed, the relational schema is logically independent of the conceptual model, as long as it preserves the semantics captured in the higher-level model.
This logical independence creates an opportunity: rather than using a fixed rule-based translation, we can \emph{optimize} the relational schema with respect to the expected workload. This allows us to
exploit the fact that different relational realizations of the same conceptual model can vary substantially in join complexity, redundancy, update cost, and indexability. In fact, there is no reason for
the underlying relational schema to remain fixed; through continuous observation and tuning, it can adapt to the changing data and workload characteristics.

\begin{figure}
  \centering
  \vspace{0pt}
  \includegraphics[scale=0.33,
  trim=3.5cm 7.7cm 0cm 3.5cm,
  clip]{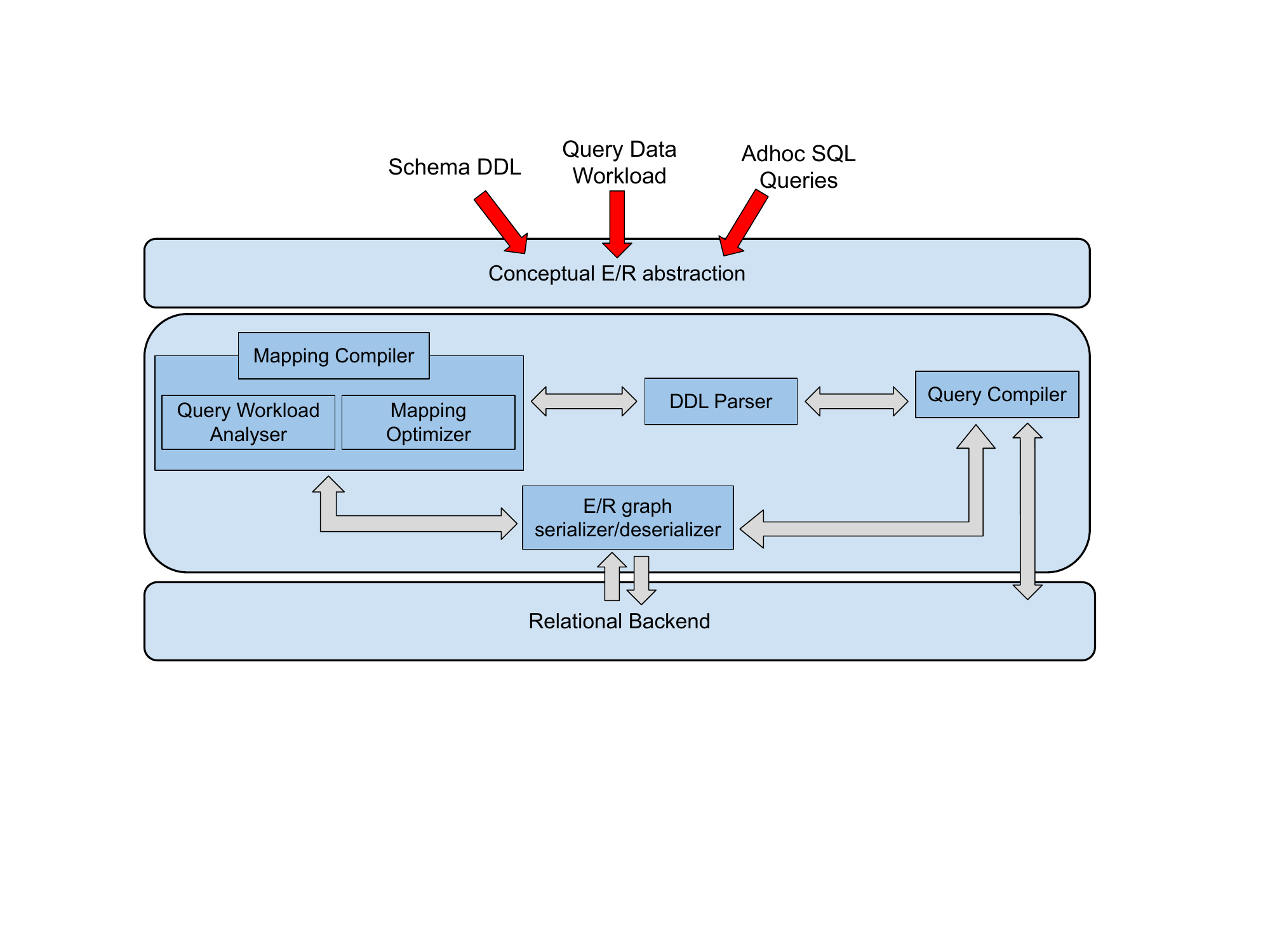}
  \caption{ErbiumDB High-level Architecture (Section~\ref{sec:implementation})}
  \vspace{-14pt}
  \label{fig:erbiumdb}
\end{figure}

\begin{figure*}[t]
  \centering
  \includegraphics[scale=0.45,
  trim=0.1cm 9.5cm 0.5cm 3.5cm,
  clip]{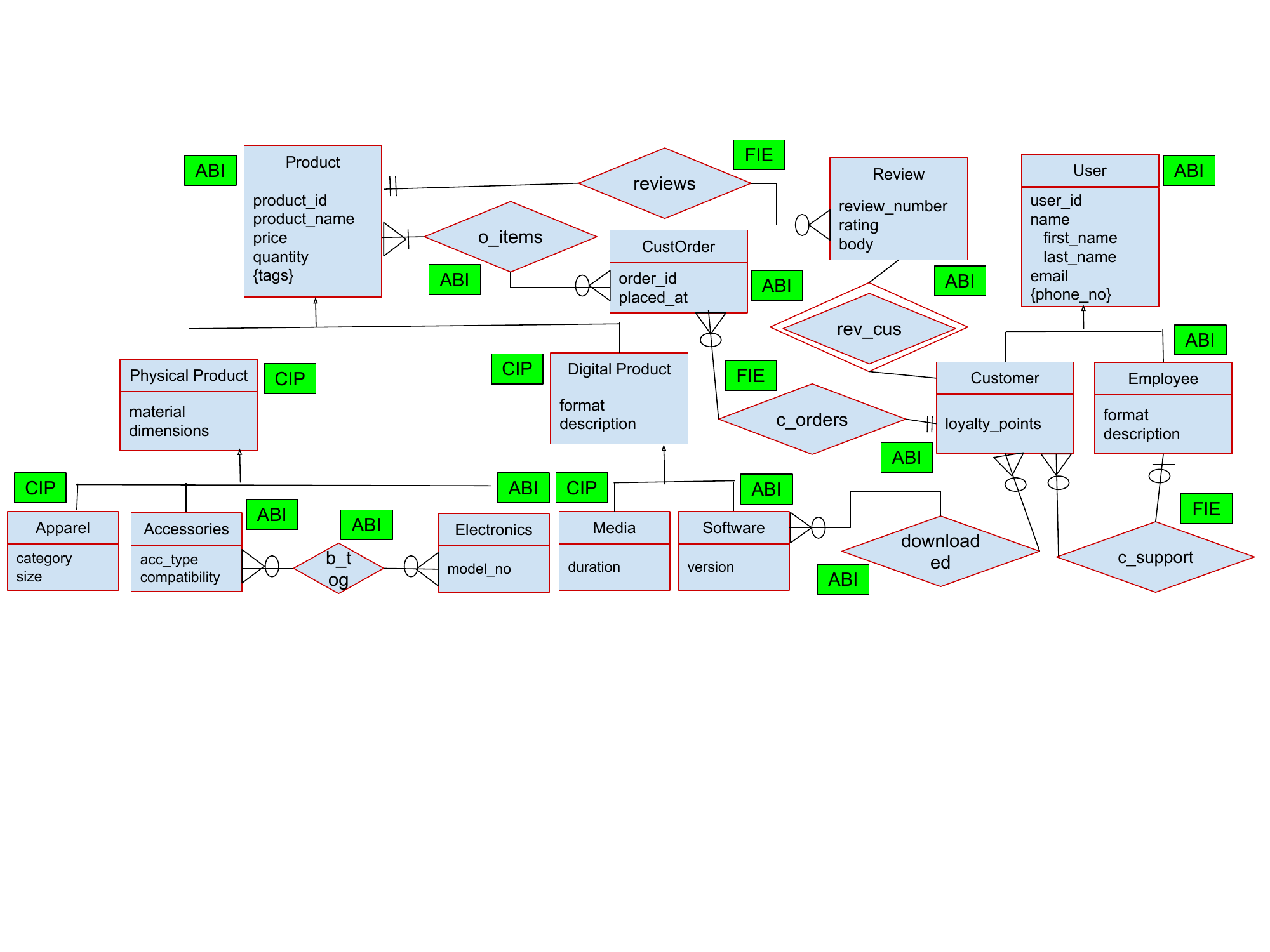}
  \vspace{-2pt}
  \caption{Example of an E/R schema of a Retail platform. Green boxes indicate mapping option chosen for each E/R component except for multi-valued attributes (mapping options defined in Section~\ref{subsec:design_variables}).
  }
  \label{fig:running_example}
\end{figure*}

In this paper, we develop a systematic framework for \emph{workload-aware optimization} of the mapping from an E/R model to a relational schema. We adopt the E/R model as the higher-level abstraction natively supported by the database system because of its widespread use in the conceptual schema design phase, its similarity to ORMs, and its ability to generalize to other data models, such as property graphs.
Our prototype, ``ErbiumDB", as shown in Figure~\ref{fig:erbiumdb}, adopts the E/R model as its primary conceptual abstraction and provides an SQL-based query interface supporting standard
SELECT and INSERT operations against the entity and relationship sets. A dedicated DDL is introduced to allow users to specify the E/R schema directly. 
ErbiumDB is built on top of PostgreSQL, which serves as the seamlessly integrated underlying relational backend. The system incorporates a cost model that automatically determines the most efficient
relational schema for a given workload. 
The query wrapper transparently converts a subset of SELECT and INSERT queries written against the conceptual schema into equivalent queries on the optimized relational
schema. 

\noindent
Our key contributions can be summarized as follows:
\begin{itemize}
    \item Formal characterization of the solution space of semantically valid relational schemas derived from a given E/R model.
    \item Workload-aware cost model that captures the primary performance trade-offs across alternative mappings.
    \item An optimization algorithm for finding effective mappings.
    \item Experimental evaluation demonstrating substantial performance improvements over rule-based mappings.
\end{itemize}

We begin with a brief discussion of relevant background, including commonly used mapping techniques (Section~\ref{sec:background}), followed by a motivating experiment that demonstrates how mapping performance varies across workloads (Section~\ref{sec:motivating-experiment}). We then present a formalization
of the solution space (Section~\ref{sec:solution-space}), and develop a cost model to compare mappings (Section~\ref{sec:cost-model}). We then present a
greedy algorithm that we use for our experimental evaluation (Section~\ref{sec:algorithm}), briefly discuss some implementation details (Section~\ref{sec:implementation}), and present a comprehensive experimental
evaluation (Section~\ref{sec:experiments}).
\vspace{0pt}

\section{Background}
\label{sec:background}

Database design is a multi-step process that begins with gathering business requirements and translating them into progressively more detailed levels of design. 
In the conceptual design phase, these requirements are modeled using a high-level data model, most notably the Entity-Relationship (E/R) model, which represents data through entities (objects of interest),
attributes (properties of entities), and relationships (associations among entities). As applications grew more complex, the model evolved into the Enhanced E/R (EER) model, incorporating
specialization, generalization, and inheritance. In the logical design phase, the conceptual schema is mapped onto the implementation data model supported by the database system, most
commonly the relational model, resulting in a relational schema. Finally, in the physical design phase, system-specific implementation decisions such as indexing, storage structures, and
views are made to optimize performance~\cite{silberschatz2019database}.

\begin{figure*}[t]
  \centering
  \vspace{-4pt}
  \includegraphics[scale=0.35,
  trim=0cm 0cm 0cm 0cm,
  clip]{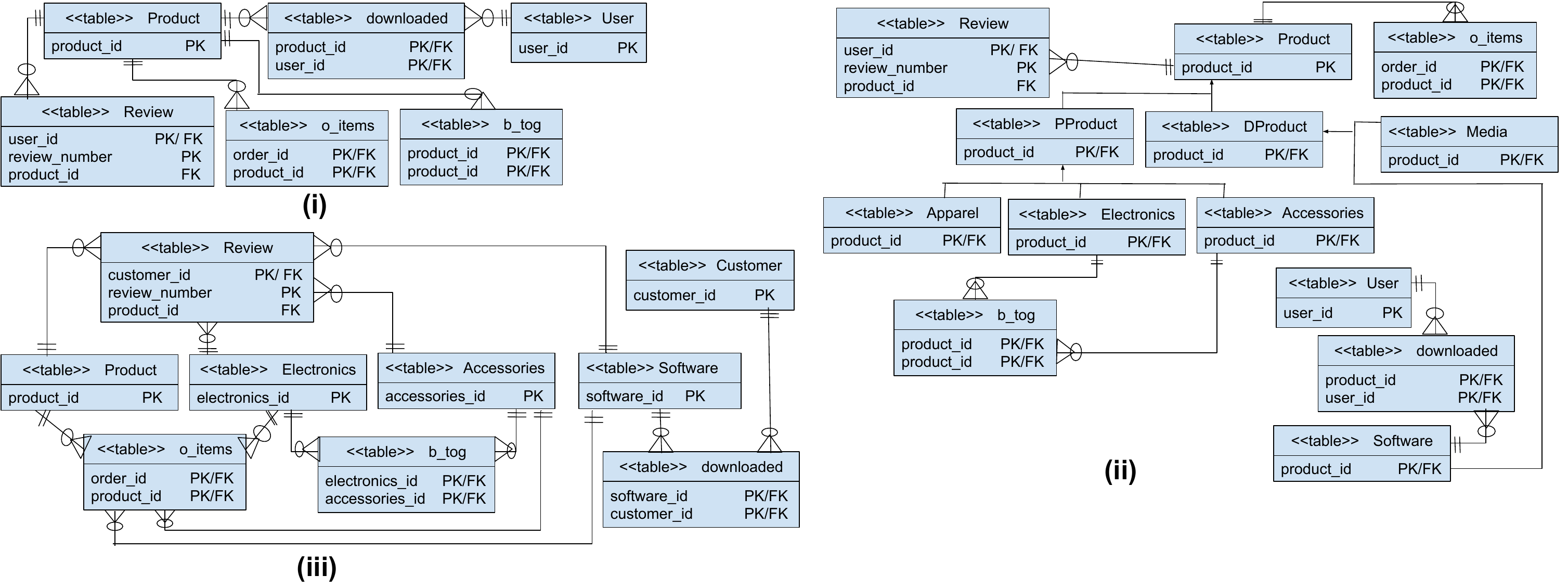}
  \vspace{-10pt}
  \caption{Three alternative relational mappings for the running example.
(i) Single-table inheritance for Product and User.
(ii) Same as (i), except Product uses class-table inheritance.
(iii) Hybrid Product mapping with selected subclasses stored separately,
and Customer mapped to a separate table.
Only primary-key and foreign-key attributes are shown.}
  \vspace*{0em}
  \label{fig:running_example_solutions}
\end{figure*}

\vspace{-1pt}
\subsection{E/R Constructs and Standard Mapping Techniques}

\noindent\textbf{Strong entity sets.}
An entity set groups entities with the same attributes.
Strong entities have their own keys, such as
\texttt{product\_id} for \texttt{Product}
(Figure~\ref{fig:running_example}). Outside an inheritance
hierarchy, each strong entity set maps to one table.\\

\noindent\textbf{Weak entity sets.}
A weak entity is identified by its owner's key and a
partial key. For example, a \texttt{PaymentMethod} is
identified by its \texttt{Customer}'s key and
\texttt{paymethod\_id}. A weak entity set can be stored in
a separate table or folded into its owner as an array of
composite values.\\

\noindent\textbf{Relationships.}
Relationship sets associate entities and may carry
attributes and participation constraints. An M:N
relationship typically maps to a separate table containing
the participant keys and relationship attributes. A 1:N
or 1:1 relationship can use a separate table or be folded
into one endpoint's table through a foreign key. For 1:N
relationships, the foreign key goes on the many side.
For 1:1 relationships, either endpoint can store it.\\

\noindent\textbf{Class Hierarchies:} Subclasses inherit attributes and relationships from their
superclasses and may add attributes. Disjointness constraints
govern membership in multiple subclasses. There are three classical strategies for mapping an inheritance hierarchy from the conceptual design to the relational model:
\begin{itemize}[itemindent=1pt, left=2pt]
    \item \textbf{Table-per-hierarchy (single-table inheritance):} All entity sets in the hierarchy are stored in a single table, with a discriminator column identifying the specific entity type represented by each row.
    
    \item \textbf{Table-per-type (class-table inheritance):} Each entity set is mapped to a separate table. The superclass table stores common attributes, while each subclass table stores only subclass-specific attributes and references the superclass table via a shared primary key.
    
    \item \textbf{Table-per-concrete-class (concrete-table inheritance):} Each concrete (instantiable) subclass is mapped to a separate table containing all inherited and subclass-specific attributes; abstract superclasses are not materialized as tables.
\end{itemize}

\noindent\textbf{Composite and multivalued attributes.}
A composite attribute groups subattributes, such as street
and city in an address. A multivalued attribute holds
multiple values per entity or relationship, such as a
user's phone numbers. It can be stored in a separate
table or as an array in the owning table.

\subsection{Running Example}
\label{subsec:running_example}

Consider the E/R diagram of a Retail platform (Figure \ref{fig:running_example}), and 
three valid alternative conceptual-to-relational mappings for it. 

\textbf{Alternative mapping~1} (partial schema shown in Figure~\ref{fig:running_example_solutions}(i)) maps the inheritance hierarchy rooted at Product into a single table (single-table inheritance), and similarly maps the hierarchy rooted at User into a single table. Each
non-hierarchical entity set is mapped to its own table. Many-to-one relationship sets are folded into the many-side table, many-to-many relationship sets are mapped to separate relationship tables, and
each multi-valued attribute is represented as an array column in the corresponding entity/relationship table. \textbf{Alternative mapping~2} (Figure~\ref{fig:running_example_solutions}(ii)) is identical to Alternative~1 except that the inheritance
hierarchy rooted at Product is mapped using class-table inheritance (one table for the superclass and one per subclass, joined by keys). \textbf{Alternative mapping~3} (Figure~\ref{fig:running_example_solutions}(iii)) is identical to Alternative~1, except that the Product hierarchy uses a modified form of single-table inheritance and the User hierarchy uses a modified form of concrete-table inheritance. In the Product
hierarchy, all descendant entity sets of Product, except Accessories, Electronics, and Software, are mapped to a single table, while the latter are each mapped to separate tables containing all their attributes, including inherited ones, and storing tuples exclusively for that class. In the User hierarchy, each class (including the superclass) is mapped to a separate table containing all its attributes, including
inherited ones, and each such table stores tuples exclusively for that class.

\textbf{These alternatives can lead to substantially different query plans, with widely varying costs.} For example, to retrieve the full set of \texttt{Product} attributes (i.e., attributes defined for the \texttt{Product} entity
        set) for each product that has been purchased in at least one \texttt{CustOrder}, Alternatives~1 and~2 can join \texttt{o\_items} with a single (logical) \texttt{Product} table. In
contrast, Alternative~3 requires joining \texttt{o\_items} with each table corresponding to \texttt{Accessories}, \texttt{Electronics}, \texttt{Software}, and \texttt{Product} and its remaining
descendants, followed by a union of the results, since product tuples are partitioned across multiple class-specific tables. 

On the other hand, consider a query that retrieves all information about
\texttt{Software} products with \texttt{download\_count} $> 100$. If $n_{\texttt{Software}}$ (i.e., the number of entities which belong to Software entity set) is small relative to the total number of product tuples across all subclasses, Alternative~3 can be substantially more efficient, as it accesses only the \texttt{Software} table. In contrast, Alternatives~1 and~2 may incur higher costs because they must scan or filter a larger product representation (e.g., a wide single table in Alternative~1 or joins across superclass and subclass tables in Alternative~2), even though only a small subclass is relevanrelationalt. Similarly, for a query that retrieves full attribute sets of \texttt{Accessories} and \texttt{Electronics} that are frequently bought together (e.g., via \texttt{b\_tog}), Alternative~3 is advantageous, as it can directly access the corresponding subclass tables without scanning all \texttt{Product} tuples.

\textbf{Schema choices also affect query complexity for relationship traversal.} For instance, to retrieve full information about a \texttt{Review} along with the reviewed \texttt{Product}, Alternatives 1 and 2 yield relatively straightforward queries because \texttt{Reviews} can reference a single product representation. In Alternative~3, however, the \texttt{Product} representation is distributed across three tables, requiring the construction of a union view before joining with \texttt{Review}, thereby complicating query generation. This complexity can increase further if the \texttt{User} hierarchy is also distributed across multiple tables, since assembling complete information for a review (i.e., \texttt{review(customer\_id, review\_number)} along with product and customer full context) may require additional joins and unions, as \texttt{Review} is a weak entity dependent on \texttt{Customer}. Overall, this example illustrates that conceptual-to-relational compilation admits many non-obvious yet valid relational schemas. This motivates (i) a systematic, workload-aware approach for selecting mapping options, and (ii) a principled mechanism for bidirectional query translation between the E/R abstraction and its relational realizations.

\section{Motivating Experiment}
\label{sec:motivating-experiment}

To illustrate the impact of logical design choices on workload performance, we conducted an end-to-end experiment using a synthetic Retail E/R schema which is extended from that is shown in Figure~\ref{fig:running_example}.
The extended schema contains 47 entity sets: 34 strong entity sets and 13 weak entity sets. Among the strong entity sets, 26 form two inheritance hierarchies. The \texttt{Product} hierarchy contains
21 entity sets across five levels, and the \texttt{User}
hierarchy contains five entity sets across three levels. 
The schema also contains 20 relationship sets: three recursive unary relationships and 17 binary relationships. Seven relationship sets have 1:N cardinality, and the remaining 13 have M:N cardinality. The data generation approach is explained in Section~\ref{subsec:data_gen}. The database size is approximately 11 GB. 

We constructed ten workloads over the E/R schema, each comprising 100 complex select queries and a corresponding set of insert operations on entity and relationship sets. The select queries use selection, projection, joins, and aggregation, and may access one or more entity or relationship
sets. Predicate selectivities range from 0.1 to 0.9. The workloads vary in their mix of high-, medium-, and low-selectivity predicates and in the hierarchy levels they access. Some focus on leaf subclasses, others mix leaf and intermediate subclasses, and others emphasize classes higher in the hierarchies. Section~\ref{subsec:query_example} presents an example select query.

We evaluate each workload across six fixed relational configurations obtained by combining three inheritance mappings with two mappings for 1:N relationships. For inheritance hierarchies, we consider:
(i) a modified form of concrete-table inheritance, in which every class, including the superclass, is mapped to a separate table containing all inherited and locally defined attributes, with each table storing only entities belonging directly to that class;
(ii) class-table inheritance; and
(iii) single-table inheritance.
For each inheritance mapping, 1:N relationships are either folded into the many-side table or stored in separate relationship tables, yielding six configurations. 
Table~\ref{tab:fixed-configurations} lists the six resulting
configurations, labeled A--F.
\begin{table}[h]
\vspace{0pt}
\centering
\caption{Six fixed relational configurations.}
\label{tab:fixed-configurations}
\small
\setlength{\tabcolsep}{4pt}
\begin{tabular}{@{}lcc@{}}
\toprule
\textbf{Inheritance mapping} &
\textbf{1:N folded} &
\textbf{1:N separate} \\
\midrule
Modified concrete-table & A & B \\
Class-table             & C & D \\
Single-table            & E & F \\
\bottomrule
\end{tabular}
\end{table}
Across all configurations, strong entity sets outside the inheritance hierarchies, weak entity sets, and M:N relationship sets are mapped to separate tables. Multivalued attributes of entities and relationships are stored as array columns in their corresponding tables. In every configuration, we create only primary-key indexes on the generated tables.

\begin{figure}[h]
  \centering
  \includegraphics[scale=0.4,
  trim=0.3cm 1cm 0cm 0.5cm,
  clip]{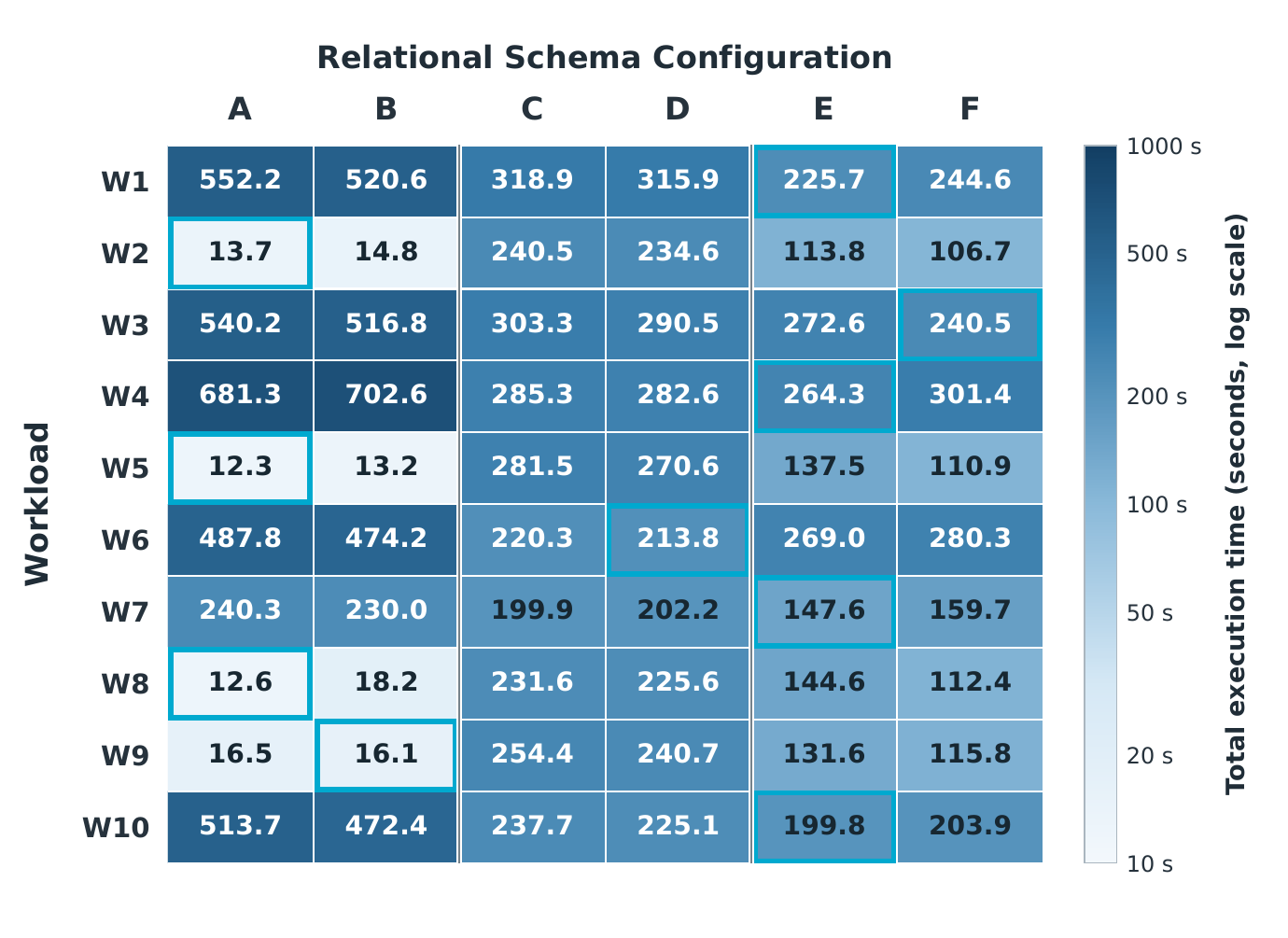}
  \caption{Total PostgreSQL execution time (in seconds) per workload for six equivalent relational schemas.}
  \label{fig:heatmap-ecommerce-er-complex-6-configs}
  \vspace{-9pt}
\end{figure}

Figure~\ref{fig:heatmap-ecommerce-er-complex-6-configs} visualizes the total PostgreSQL execution time (in seconds) for the ten complex E/R query workloads. For each mapping, we rewrite every complex query expressed over the E/R schema into an equivalent relational
query over the relational schema produced by that mapping. We execute the rewritten queries in PostgreSQL and report the total execution time for each workload.

The results demonstrate that logical design choices substantially affect workload performance, with no single configuration performing the best across all workloads. For W5, execution time ranges from 12.3 seconds under Config~A to 281.5 seconds under Config~C, a $22.9\times$ difference. However, Config~A is substantially less efficient on W1, requiring 552.2 seconds compared with 225.7 seconds
under Config~E. Thus, modified concrete-table inheritance is preferable for some workloads, while single-table inheritance is preferable for others. Class-table inheritance also achieves the lowest execution time on W6 through Config~D, showing that each of the three inheritance strategies can be advantageous.
Relationship mapping choices introduce additional workload-dependent tradeoffs. Even with the inheritance mapping held fixed, folding 1:N relationships is not uniformly beneficial: Config~E, which folds these relationships, outperforms Config~F on W1, whereas Config~F is faster on W3. 

These results clearly demonstrate the importance of workload-aware approach to logical design tuning.

\section{Formalizing the Solution Space}
\label{sec:solution-space}
Next, we discuss how to formally characterize the space of semantically valid relational schemas that can be derived from a fixed E/R model, so we can systematically reason 
through the performance implications.

\begin{figure*}
  \centering
  \includegraphics[scale=0.61,
  trim=1.6cm 13.5cm 0.5cm 0.5cm,
  clip]{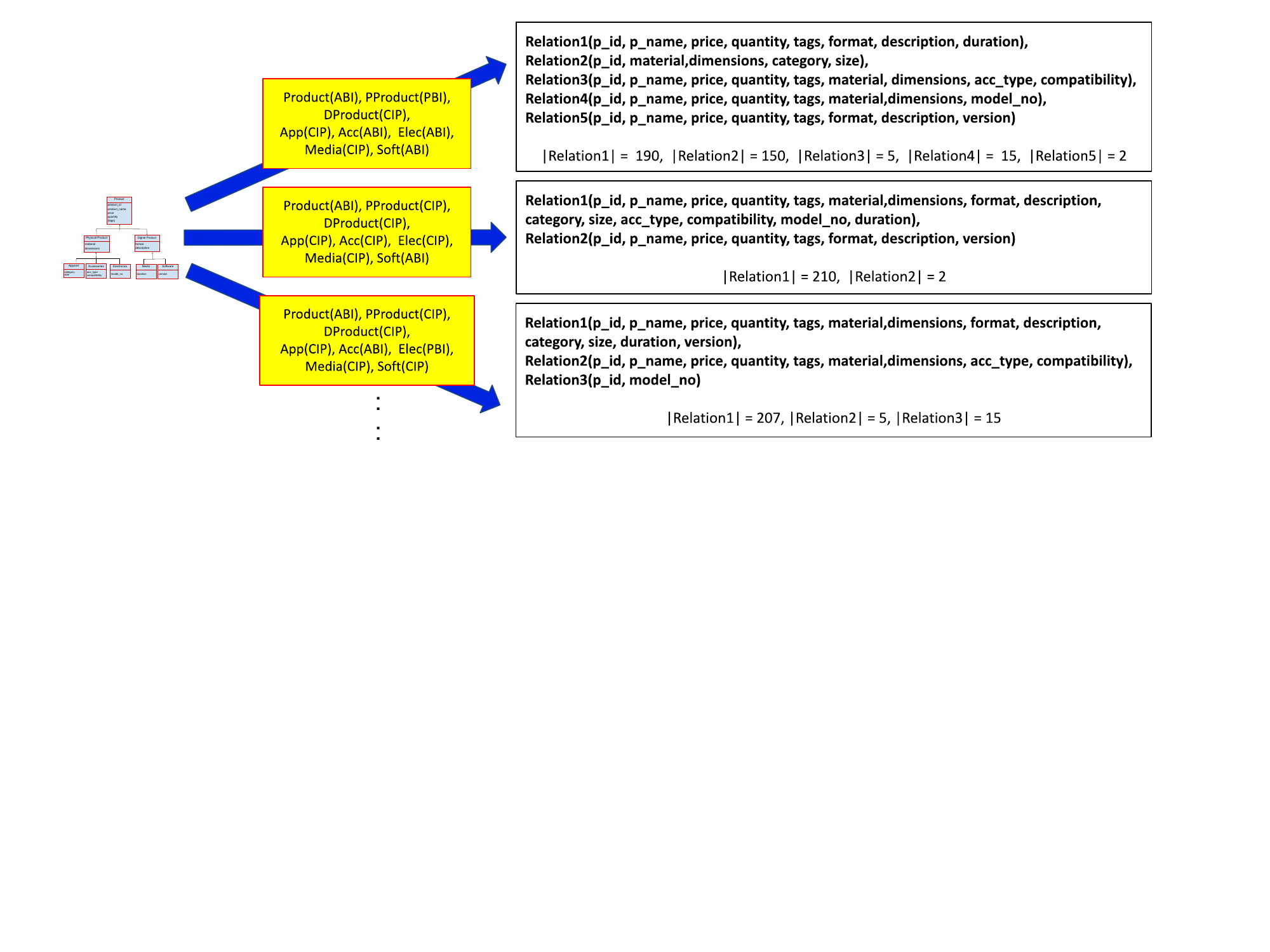}
 \caption{Three (out of $3^7$) different mapping options for the Product hierarchy. The tuple cardinality of each most specific subtype is 
$\lvert\mathrm{Product}\rvert=10$,
$\lvert\mathrm{PProduct}\rvert=50$,
$\lvert\mathrm{DProduct}\rvert=20$,
$\lvert\mathrm{App}\rvert=100$,
$\lvert\mathrm{Acc}\rvert=5$,
$\lvert\mathrm{Elec}\rvert=15$,
$\lvert\mathrm{Media}\rvert=10$, and
$\lvert\mathrm{Soft}\rvert=2$.}
 \vspace*{0em}
  \label{fig:running_example-hierarchy}
\end{figure*}

\subsection{Design Variables}
\label{subsec:design_variables}

We introduce discrete decision variables that capture mapping choices for each component of the E/R diagram, and determine how conceptual constructs are translated into relational structures.

\paragraph{Strong (Regular) Entity Sets.}
A strong entity set that does not inherit from another entity set  is mapped using the \textbf{All By Itself (ABI)} option (i.e., there is no choice), which generates a separate table containing all attributes of the entity
(with the possible exception of multi-valued attributes as discussed below).
\paragraph{Weak Entity Sets.}
A weak entity set can be mapped using either: (i) \textbf{All By Itself (ABI)}, which creates a separate table containing all its attributes (plus the primary key of owner entity), or (ii) \textbf{Contained In Parent (CIP)}, which merges the weak entity into its identifying entity as an array of composite type. For example, consider a weak entity 
\texttt{Review(\dotuline{review\_number}, rating, body)} that depends on \texttt{Customer}, capturing the reviews made by a customer on products. 
Each review is identified by the composite key \texttt{(customer\_id, review\_number)}. Under the CIP option, a composite type (e.g., \texttt{review\_type}) is defined, and the \texttt{Customer} table is extended with an array column of this type. The reviews associated with each customer are stored as elements of the array and retrieved using \texttt{UNNEST}. 

\paragraph{Class Hierarchies.}
Mapping choices for class hierarchies correspond to different inheritance strategies, as seen in the Product and User hierarchies in the running example.
The root superclass is always mapped using \textbf{ABI}, generating a table for itself. The table contains all attributes of itself, and may also include attributes from descendant subclasses through
upward insert propagation (i.e., an insert into a descendant entity set results in an insert into the table). For each subclass, three options are available:\\ (i) \textbf{ABI}, which creates a table for that entity set. The table contains all attributes of the subclass, including
all inherited attributes, and may also include attributes from descendant subclasses through upward insert propagation. Inserts into this entity set do not propagate upward in the hierarchy;\\ (ii)
\textbf{Partially By Itself (PBI)}, which also generates a table for the entity set, but the table contains only the attributes defined at that subclass level, and may also include attributes from descendant
classes through upward insert propagation. Inserts propagate upward in the hierarchy; and\\ (iii) \textbf{CIP}, which does not generate a separate table for subclass. The subclass is merged into its parent relation. Inserts propagate upward in the hierarchy. 

\paragraph{Relationship Sets.}
Each relationship set (unary or binary) can be mapped using either: (i) \textbf{All By Itself (ABI)}, which creates a separate relationship table, or (ii) \textbf{Folded In Entity (FIE)}, which embeds the relationship by introducing a foreign key into one participating entity. We consider only \textbf{ABI} option for many-to-many relationships.

\paragraph{Multi-valued Attributes.}
Each multi-valued attribute can be mapped using either: (i) \textbf{All By Itself (ABI)}, which creates a separate table, or (ii) \textbf{Contained In Parent (CIP)}, which stores the attribute as an array within the parent entity.

\vspace{5pt}
Together, these design variables define a combinatorial solution space, where each assignment corresponds to a valid relational schema, provided that it satisfies the validity constraints defined in
Section~\ref{subsec:validity_constraints}. For instance, the mapping choices corresponding to Solution~3 in Section~\ref{subsec:running_example} are highlighted by green boxes in
Figure~\ref{fig:running_example}. To illustrate the diversity of relational schema configurations that can arise from different mappings, Figure~\ref{fig:running_example-hierarchy} presents several
relational schema instances, along with their corresponding table cardinalities, restricted to the \texttt{Product} hierarchy.

\subsection{Validity Constraints}
\label{subsec:validity_constraints}

A complete mapping is obtained by selecting one applicable option for each component in the E/R schema. At first glance, one might expect these choices to be incompatible in many cases. Somewhat surprisingly, this is largely not the case. In particular, within a class hierarchy, the mapping choice for each non-root entity set can be made independently of the choices for the other entity sets in that hierarchy.

\begin{lemma}[Independence of Mapping Choices in a Class Hierarchy]
\label{lem:hierarchy-validity}
Consider a class hierarchy with $n+1$ entity sets: one root entity set and $n$ non-root entity sets at arbitrary levels of the hierarchy. Except for the root, which must use the ABI option, every
assignment of options from $\{\text{ABI}, \text{PBI}, \text{CIP}\}$ to the $n$ non-root entity sets yields a valid mapping configuration, resulting in $3^n$ valid mapping configurations.
\end{lemma}

The only cross-component validity constraints arise from weak entity sets. First, if a weak entity set participates in a relationship set, then mapping that weak entity set using CIP is invalid 
(since the relationship tuples will not be able to refer to a foreign key). Second, if a weak entity set is mapped using CIP, then every multi-valued attribute belonging to that weak entity set must also be mapped using CIP; mapping such an attribute using ABI is invalid.

Outside of these two exceptions, the remaining choices for weak entity sets, relationship sets, and multi-valued attributes are independent. Thus, if $\mathcal{H}$ is the set of class hierarchies, $n_H$
is the number of non-root entity sets in hierarchy $H$, $\mathcal{W}$ is the set of weak entity sets, $\mathcal{R}$ is the set of relationship sets, and $\mathcal{M}$ is the set of multi-valued
attribute sets, then ignoring the two weak-entity exceptions and assuming many-to-one relationships (which have two choices: ABI and FIE), the number of candidate mappings is roughly: 
$N_0 \;=\; \left( \prod_{H \in \mathcal{H}} 3^{n_H} \right) \cdot 2^{|\mathcal{W}| + |\mathcal{R}| + |\mathcal{M}|}$.

\subsection{Extensions}

In the mapping space studied here, we represent many-to-many relationships as separate tables. Another possible mapping folds the relationship into a participating entity as a collection of structured values (similar to \textbf{CIP} for weak entities). Our approach can be extended to support this and other denormalization choices by adding options to the relevant mapping variables. Each new option requires corresponding rules for validity, query compilation, and cost estimation.
Multiple inheritance can also be accommodated through
a preprocessing step inspired by hypertree decomposition.
Regions that induce cycles in the underlying undirected
inheritance graph are contracted into composite nodes
with union schemas, using NULLs for attributes that are
inapplicable to a particular subtype. We omit the details
for brevity.

\section{Workload-Aware Cost Model}
\label{sec:cost-model}
To compare two candidate relational schemas for the same E/R diagram, ideally we would materialize each schema and populate with representative data, and execute the full workload on each to measure and compare the total execution times. 
This is especially the case because, as observed in the seminal work on \emph{auto-tuning}~\cite{autoadmin}, the query optimizer behavior in response to physical schema changes is hard to model. However, this exhaustive approach is computationally 
infeasible given the exponential size of the design space and also the complex queries that get generated by the mapping process. 

In this initial work, we instead focus on developing a simpler and easier-to-use cost model that is effective at comparing candidate mappings. We observe that the differences across candidate relational schemas start showing up for 
relatively simple queries, namely scans of entities and relationships (i.e., \texttt{SELECT * FROM $E$} and \texttt{SELECT * FROM $R$} at the conceptual layer). As discussed earlier, the costs of these simple operations can be wildly 
different for different physical schemas, and we argue that it is sufficient to focus on these queries for identifying the right physical schema. 

Next, we discuss how we map an arbitrary workload to a workload with just these scan queries (with appropriate weights), and then discuss how to model the costs of these scans for different candidate physical schemas.
\vspace{-3pt}

\subsection{Workload Model}

We assume a representative workload expressed as a set of query templates, a corresponding weight vector that captures their frequencies, and relevant data statistics such as insert frequencies, including cardinalities of entity and relationship sets.

Let $\mathcal{C}$ denote the set of entity sets,
relationship sets ($\mathcal{R}$), and multivalued attribute sets ($\mathcal{M}$) in the
input E/R schema. Let $\mathcal{W_I}$ denote the set of insert queries in the workload $\mathcal{W}$ defined over the E/R schema. For each
component $c_i \in \mathcal{C}$, let $\mathcal{I}_i$
denote the total number of insert operations in
$\mathcal{W}_I$ that target $c_i$.

Let $\mathcal{W_S}$ denote the set of select queries in the workload $\mathcal{W}$ defined over the E/R schema. We define a set of query templates as follows. Each template corresponds to a
\texttt{SELECT * } query over an entity set or relationship set. For a non-hierarchical strong entity set, the query retrieves the complete set of attributes defined for that entity set. For a
hierarchical strong entity set, the query retrieves both the attributes defined for the entity set itself and all attributes inherited from its ancestor entity sets. 
For a weak entity set, the query retrieves all of its attributes together with the attributes of its identifying (owning) entity sets.
If the identifying strong parent participates in an inheritance hierarchy, its complete attribute set consisting of both its own attributes and all inherited attributes, is included. For a relationship
set, the query retrieves all attributes defined for the relationship together with the complete attribute sets of each participating entity set. 

Let $\mathcal{T} = \{T_1, T_2, \dots, T_n\}$ denote the set of \texttt{SELECT * }  query templates defined over the entity and relationship sets of the E/R model abstraction where $n$ is the total number of entity and relationship sets in input E/R schema. 
Each query in $\mathcal{W}_S$ is decomposed into one or more canonical entity and relationship templates. A query maps to every template corresponding to an entity or relationship whose complete conceptual state is required to answer the query. The weight of a template is the aggregate frequency of all queries that map to that template.
We define a weight vector
\[
\mathbf{w}=(w_1,w_2,\ldots,w_n),
\]
where $w_i$ denotes the aggregate frequency of template $T_i$ in the workload. Formally,
\[
w_i =  \sum_{q\in\mathcal{W}_S:, q\mapsto T_i}
f(q),
\]
where $f(q)$ is the execution frequency of query $q$ in the workload, and $q\mapsto T_i$ indicates that query $q$ maps to template $T_i$. Consequently, the weight vector $\mathbf{w}$ summarizes the workload demand on each canonical entity and relationship template.

In practice, the workload parameters can be estimated from standard sources such as the query logs and ORM-generated query traces; these can be clustered into templates and aggregated to estimate query
and insert frequencies for individual entity and relationship sets. 
Under our structural workload abstraction, a multi-way join contributes access requests to each participating entity and relationship set, allowing its contribution to be summarized by the corresponding canonical requests.
Finally, database statistics catalogs can provide approximate cardinalities and selectivities for the relevant entity, relationship, and attribute sets. We do not develop this estimation procedure in the present work, and instead assume that such workload and catalog information is available.

\subsection{Cost Estimation}
\label{subsec:cost_estimation}

Let $S_{valid}$ denote the subset of valid mapping configurations (that satisfy the validity constraints described in Section~\ref{subsec:validity_constraints}). 
The total cost of a valid schema $S \in \mathcal{S}_{\text{valid}}$ is then defined as
\[
C_{\text{total}}(S)
=
\sum_{T_i \in \mathcal{T}} w_i \cdot C(T_i, S)
+
\sum_{j} I_j \cdot C_{\text{ins}}(j,S),
\]

where $\mathcal{T}$ is the set of query templates, $w_i$ is the frequency associated with template $T_i$ in the workload, $C(T_i,S)$ represents the estimated execution cost of query template $T_i$ under schema $S$, $I_j$ is the insert workload count of component $j$, and $C_{\text{ins}}(j,S)$ denotes the estimated cost of an insert operation targeting component $j$ under the same schema. For each query template $T_i$ defined over the E/R model abstraction, we generate the corresponding query over the candidate relational schema $S$ and estimate $C(T_i,S)$ using the translated query. 
The optimal relational design is therefore obtained as
\[
S^{*} = \arg\min_{S \in \mathcal{S}_{\text{valid}}} C_{\text{total}}(S).
\]

\subsubsection{SELECT * Cost}

Given a \texttt{SELECT *} query template over an entity or relationship set, we estimate its cost by decomposing the logical access into
the physical operations required under the chosen mapping configuration. We don't try to model every execution
detail of the backend, but rather try to capture the dominant structural effects of mapping choices, namely the 
additional scans, joins, unions, and expansion operations needed to reconstruct the logical tuple.

Table~\ref{tab:materialization-costs} summarizes the cost formulas for \texttt{SELECT *} query templates of entities and relationships. 

\begin{table*}[t]
\centering
\caption{Structural materialization cost rules for \texttt{SELECT *} queries, omitting orthogonal multi-valued attribute retrieval costs.}
\label{tab:materialization-costs}
\scriptsize
\setlength{\tabcolsep}{5pt}
\renewcommand{\arraystretch}{1.45}

\begin{tabular}{@{}
>{\raggedright\arraybackslash}p{0.20\textwidth}
!{\vrule width 0.3pt}
>{\raggedright\arraybackslash}p{0.76\textwidth}
@{}}
\toprule
\textbf{Case} & \textbf{SELECT * Query Cost} \\
\midrule

Regular entity $E$, $M(E)=\mathrm{ABI}$
&
\(\displaystyle
\mathsf{Scan}(F_M(E))
\)
\\
\midrule

Hierarchy Entity $E$, $M(E)=\mathrm{ABI}$
&
\(\displaystyle
\mathsf{Scan}(F_M(E))
+
\mathbf{1}\!\left[\mathsf{ABIDesc}_M(E)\neq\emptyset\right]
\mathsf{Union}(F_M(E))
+
\sum_{D\in\mathsf{ABIDesc}_M(E)}
\left[
\mathsf{Scan}(F_M(D))
+
\mathsf{Union}(F_M(D))
\right]
\)
\\
\midrule

Hierarchy Entity $E$, $M(E)=\mathrm{PBI}$
&
\(\displaystyle
\sum_{A\in\mathsf{AncMat}_M(E)}
\mathsf{INLJ}(F_M(E),F_M(A))
+
\mathbf{1}\!\left[\mathsf{ABIDesc}_M(E)\neq\emptyset\right]
\left(
\mathsf{Scan}(F_M(E))
+
\mathsf{Union}(F_M(E))
\right)
+
\sum_{D\in\mathsf{ABIDesc}_M(E)}
\left[
\mathsf{Scan}(F_M(D))
+
\mathsf{Union}(F_M(D))
\right]
\)
\\
\midrule

Hierarchy Entity $E$, $M(E)=\mathrm{CIP}$
&
\(\displaystyle
\begin{aligned}
&\mathsf{Scan}(F_M(a_M(E)))
+
\sum_{A\in\mathsf{AncMat}_M(a_M(E))}
\mathsf{INLJ}(F_M(E),F_M(A))
\\
&\quad+
\mathbf{1}\!\left[\mathsf{ABIDesc}_M(E)\neq\emptyset\right]
\left(
\mathsf{Scan}(F_M(E))
+
\mathsf{Union}(F_M(E))
\right)
\\
&\quad+
\sum_{D\in\mathsf{ABIDesc}_M(E)}
\left[
\mathsf{Scan}(F_M(D))
+
\mathsf{Union}(F_M(D))
\right]
\end{aligned}
\)
\\
\midrule

Weak entity $W$, $M(W)=\mathrm{ABI}$
&
\(\displaystyle
\sum_{B\in\mathcal{B}_O}
\left[
S_B^{\mathrm{CIP}}
+
\mathsf{INLJ}(N_W,F_M(B))
+
\sum_{A\in\mathsf{AncMat}_M(B)}
\mathsf{INLJ}(N_W,F_M(A))
\right]
\)
\\
\midrule

Weak entity $W$, $M(W)=\mathrm{CIP}$
&
\(\displaystyle
\sum_{B\in\mathcal{B}_O}
\left[
S_B^{\mathrm{CIP}}
+
\sum_{A\in\mathsf{AncMat}_M(B)}
\mathsf{INLJ}(q_B^W,F_M(A))
\right]
+
\mathsf{ExpandWeak}(N_{O,W}^{+},N_W)
\)
\\
\midrule

Relationship $R$, $M(R)=\mathrm{ABI}$
&
\(\displaystyle
\sum_{B\in\mathcal{B}_{E_1}}
\left[
S_B^{\mathrm{CIP}} + P_B(N_R)
\right]
+
\sum_{B\in\mathcal{B}_{E_2}}
\left[
S_B^{\mathrm{CIP}} + P_B(N_R)
\right]
\)
\\
\midrule

Relationship $R$, $M(R)=\mathrm{FIE}$
&
\(\displaystyle
\sum_{B\in\mathcal{B}_{E_f}}
\left[
S_B^{\mathrm{FIE}}
+
\sum_{A\in\mathsf{AncMat}_M(B)}
\mathsf{INLJ}(q_B^f,F_M(A))
\right]
+
\sum_{B\in\mathcal{B}_{E_f}}
\sum_{B'\in\mathcal{B}_{E_o}} 
\left[
S_{B'}^{\mathrm{CIP}}
+
P_{B'}(q_B^f)
\right]
\)
\\

\bottomrule
\end{tabular}
\end{table*}

\paragraph{Cardinalities and folded sizes.}
For a fixed candidate configuration
$S \in \mathcal{S}_{\mathrm{valid}}$, let $M$ denote its mapping
assignment.
We first define the mapping-dependent size function \(F_M(\cdot)\), which
is used by the structural cost model. For an
entity \(E\), let \(N_E\) denote its cardinality.

For hierarchy entities, let \(\mathcal{G}=(V,\mathcal{E})\) be a rooted
inheritance tree with root \(\rho\). The root is mapped using ABI, and
each non-root node \(v\in V\) is assigned
$
M(v)\in\{\mathrm{ABI},\mathrm{PBI},\mathrm{CIP}\}.
$
Let \(b(v)\) denote the exclusive cardinality of \(v\), i.e., the number
of tuples whose most specific type is \(v\). The folded size of \(v\)
under mapping \(M\) is
\[
F_M(v)
=
b(v)
+
\sum_{c\in\mathrm{Ch}(v):\,M(c)\neq \mathrm{ABI}}
F_M(c),
\]
where \(\mathrm{Ch}(v)\) denotes the immediate children of \(v\). Thus,
an ABI child starts a separate physical branch and does not contribute to
the folded size of its parent.
If \(v\) is materialized, i.e., \(M(v)\in\{\mathrm{ABI},\mathrm{PBI}\}\),
then the generated relation has cardinality
$
|R_v|=F_M(v).
$
If \(M(v)=\mathrm{CIP}\), no separate relation \(R_v\) is generated;
instead, the \(F_M(v)\) tuples logically belonging to \(v\) are stored in
the nearest materialized ancestor relation \(R_{a_M(v)}\). For a regular non-hierarchical entity $E$, \[
F_M(E)=N_E .
\]

\paragraph{Cost primitives.}
We use a small set of structural cost primitives. For an input of
cardinality $s$,
\[
\mathsf{Scan}(s)=c_s s,
\qquad
\mathsf{Union}(s)=c_{\cup}s .
\]
For an index nested-loop join with left input cardinality $n$ and indexed
right input cardinality $s$, we use
\[
\mathsf{INLJ}(n,s)
=
n+n\log_2\max\{1,s\}.
\]

For a hierarchy node $v$, let $F_M(v)$ denote its folded size under mapping
$M$. If $v$ is materialized, then $|R_v|=F_M(v)$. If $v$ is mapped using
CIP, no relation $R_v$ is generated; instead, its tuples are stored in the
nearest materialized ancestor relation $R_{a_M(v)}$. An entity is said to be materialized if it is mapped with either ABI or PBI. 

Let $\mathsf{ABIDesc}_M(E)$ denote the ABI-mapped proper descendants of
$E$. These descendants form separate physical branches that must be
included when scanning $E$. We write
\[
\mathsf{Branches}_M(E)
=
\{E\}\cup \mathsf{ABIDesc}_M(E).
\]
Let $\mathsf{AncMat}_M(E)$ denote the materialized ancestors of $E$,
i.e., ancestors mapped using ABI or PBI, that must be probed to
reconstruct the attributes of $E$'s own component. If
$M(E)=\mathrm{ABI}$, then
$
\mathsf{AncMat}_M(E)=\emptyset,
$
since the ABI relation already contains the inherited state required for
$E$. If $M(E) = \mathrm{PBI}$, then $\mathsf{AncMat}_M(E)$ contains the
materialized ancestors on the path from $E$ to the nearest ABI ancestor,
including that ABI ancestor. If $M(E)=\mathrm{CIP}$, then $E$ is stored in the nearest materialized ancestor $a_M(E)$.
\[
\mathsf{AncMat}_M(E)=
\begin{cases}
\emptyset,
& M(E)=\mathrm{ABI},
\\[1mm]

\{a_M(E)\}\cup \mathsf{AncMat}_M(a_M(E)),
& M(E)=\mathrm{PBI},
\\[1mm]

\mathsf{AncMat}_M(a_M(E)),
& M(E)=\mathrm{CIP}.
\end{cases}
\]
where \(a_M(E)\) denotes the nearest materialized ancestor of \(E\), i.e., the nearest ancestor mapped using ABI or PBI.

For an entity $E$ mapped with CIP, define the additional cost for pruning its relevant tuples from the materialized ancestor table as,
\[
S_E^{\mathrm{CIP}} =
\begin{cases}
\mathsf{Scan}(F_M(a_M(E))), & M(E)=\mathrm{CIP},\\
0, & \text{otherwise},
\end{cases}
\]
where $a_M(E)$ is the nearest materialized ancestor of $E$.

For compactness, we omit multi-valued attribute retrieval costs from the
main tables; they are orthogonal to hierarchy reconstruction and can be
added separately when multi-valued attributes are stored in separate
relations.

\paragraph{Weak entities.}
The fifth and sixth rows in Table~\ref{tab:materialization-costs} extend the same branch-based accounting to weak
entities. For a weak entity $W$ with owner $O$, let
$
\mathcal{B}_O=\mathsf{Branches}_M(O).
$
An ABI weak entity stores its $N_W$ tuples separately and joins them to
the corresponding owner branches. For a CIP weak entity, we estimate
\[
\hspace*{-1em}
p_W=1-\left(1-\frac{1}{N_O}\right)^{N_W},
\;
q_B^W=F_M(B)p_W,
\;
N_{O,W}^{+}=N_Op_W,
\]
where $p_W$ is the probability that an owner tuple has at least
one weak-entity tuple, $q_B^W$ is the expected number of such
owner tuples in branch $B$, and $N_{O,W}^{+}$ is their expected
total number across all owner branches.
The term
\[
\mathsf{ExpandWeak}(N_{O,W}^{+},N_W)
=
\mathsf{Scan}(N_{O,W}^{+})+c_wN_W
\]
accounts for accessing those owner tuples and unfolding their embedded
weak-entity values.

\paragraph{Relationships.} 
Let $R(E_1,E_2)$ be a binary relationship with cardinality $N_R$.
For an endpoint entity $E \in \{E_1,E_2\}$, define
$
\mathcal{B}_E=\mathsf{Branches}_M(E),
$ and
$
N_E=\sum_{B\in\mathcal{B}_E}F_M(B).
$
For $q$ relationship tuple probes, define the cost of probing one endpoint branch as
\[
P_B(q)
=
\mathsf{INLJ}(q,F_M(B))
+
\sum_{A\in\mathsf{AncMat}_M(B)}
\mathsf{INLJ}(q,F_M(A)).
\]
Then the cost of joining a relationship mapped with ABI, having $N_R$ relationship tuples to endpoint entity $E$ is
\[
\sum_{B\in\mathcal{B}_E}
\left[
S_B^{\mathrm{CIP}} + P_B(N_R)
\right].
\]

For an FIE relationship folded into endpoint entity $E_f$, let $E_o$ be the
other endpoint entity. We assume $E_f$ has at most one participating relationship
tuple per entity tuple. Thus, for each branch $B\in\mathcal{B}_{E_f}$,
\[
q_B^f
=
N_R\frac{F_M(B)}{N_{E_f}}.
\]
gives the expected number of tuples in each branch which participate in relationship. The folded-side extra scan cost to prune relationship tuples for branch $B$ is
\[
S_B^{\mathrm{FIE}} =
\begin{cases}
\mathsf{Scan}(F_M(B)), & M(B)\in\{\mathrm{ABI},\mathrm{PBI}\},\\
\mathsf{Scan}(F_M(a_M(B))), & M(B)=\mathrm{CIP}.
\end{cases}
\]

\subsubsection{Insert Cost}

For each conceptual entity, relationship, or multivalued attribute node \(X\), let \(I_X\) denote its workload insert
frequency. We model the cost of inserting \(I\) tuples into a relation of
current cardinality \(s\) with \(k\) indexes as
\[
\mathsf{Ins}(I,s,k)
=
I\left(c_i+k\log_2\max\{1,s\}\right),
\]
where \(c_i\) is the base per-tuple insertion cost. Generated relations
normally have one primary-key index, so \(k=1\).

We first consider conceptual nodes other than FIE
relationships, CIP weak entities, and CIP multivalued
attributes, which require separate treatment.
For each such node $X$, 
let \(\mathcal{T}_X\) be the set
of generated relations populated by an insert into \(X\). For a
separately materialized non-hierarchy entity, weak entity, or
relationship, \(\mathcal{T}_X=\{R_X\}\). For a hierarchy entity \(E\),
\[
\hspace*{-5mm}
\mathcal{T}_E
=
\begin{cases}
\{R_E\},
& M(E)=\mathrm{ABI},
\\[1mm]
\{R_E\}\cup
\{R_A\mid A\in\mathsf{AncMat}_M(E)\},
& M(E)=\mathrm{PBI},
\\[1mm]
\{R_{a_M(E)}\}\cup
\{R_A\mid A\in\mathsf{AncMat}_M(E)\},
& M(E)=\mathrm{CIP}.
\end{cases}
\hspace*{5mm}
\]
Thus, an ABI insert affects only \(R_E\), whereas PBI and CIP inserts
populate the materialized relations needed to represent the entity state
up to the nearest ABI ancestor. The corresponding cost is
\[
I_X \cdot C_{\mathrm{ins}}(X,M)
=
\sum_{T\in\mathcal{T}_X}
\mathsf{Ins}(I_X,|T|,1).
\]

A multi-valued attribute \(A\) materialized in a separate relation
\(T_A\) contributes
\[
I_A \cdot C_{\mathrm{ins}}(A,M)
=
\mathsf{Ins}(I_A,|T_A|,1),
\]
where \(I_A\) includes contributions from hierarchy descendants when
applicable.

A CIP-mapped weak entity or FIE-mapped relationship is instead inserted
by updating its folded representation. Let \(\phi(X)\) denote the fold
target: the identifying owner for a weak entity or the folded endpoint
for a relationship. For each physical branch
\(B\in\mathsf{Branches}_M(\phi(X))\), define its update relation as
\[
\mu_M(B)
=
\begin{cases}
R_B, & M(B)\in\{\mathrm{ABI},\mathrm{PBI}\},\\[1mm]
R_{a_M(B)}, & M(B)=\mathrm{CIP}.
\end{cases}
\]
The candidate update relations are therefore
\[
\mathcal{R}_X
=
\left\{
\mu_M(B)
\mid
B\in\mathsf{Branches}_M(\phi(X))
\right\}.
\]
Because the physical branch containing the referenced tuple is not known
a priori, every relation in \(\mathcal{R}_X\) is probed, although exactly
one update succeeds. Assuming that the folded column has no separate
index, the cost is
\[
I_X \cdot 
C_{\mathrm{ins}}^{\mathrm{fold}}(X,M)
=
\sum_{T\in\mathcal{R}_X}
\mathsf{INLJ}(I_X,|T|)
+
c_i I_X .
\] 

For a fixed mapping $M$, let $\mathcal{V}_{\mathrm{fold}}$
denote weak entities mapped using CIP and relationships
mapped using FIE. Let $\mathcal{V}_{\mathrm{mva}}$ denote
multivalued attributes mapped using ABI, and let
$\mathcal{V}_{\mathrm{reg}}$ denote all remaining entity
and relationship nodes. The total workload insert cost is
\[
\begin{aligned}
\sum_{X\in\mathcal{V}_{\mathrm{reg}}}
I_X \cdot C_{\mathrm{ins}}(X,M)
+
\sum_{X\in\mathcal{V}_{\mathrm{fold}}}
I_X \cdot C_{\mathrm{ins}}^{\mathrm{fold}}(X,M)
+
\\[-1pt]
\sum_{A\in\mathcal{V}_{\mathrm{mva}}}
I_A \cdot
C_{\mathrm{ins}}(A,M).
\end{aligned}
\]

\section{Algorithms}
\label{sec:algorithm}
We develop and experiment with several variants of greedy optimization algorithms, and report results for one of them: {\bf Greedy Hill Climbing with Random Restarts}. Here, the search begins from a default schema design in
which each entity set and relationship set is mapped using the All By Itself (ABI) option, and each multi-valued attribute is mapped using the Contained In Parent (CIP) option.
The total cost of this default schema is evaluated as the initial solution. 
We define a single {\bf move} as selecting a single schema component, either an entity set or a
relationship set, and replacing its current mapping choice with the locally optimal alternative, while keeping all other components fixed. At each step, the algorithm evaluates
all possible single-component moves and chooses the one that yields the greatest reduction in cost. Since the objective is cost minimization, the algorithm performs a form of
reverse hill climbing, iteratively moving toward solutions with lower cost until no further improvement is possible. The search terminates when a local minimum is reached. 
The algorithm is executed multiple times from randomly generated feasible schema designs, in addition to starting from the default configuration. Each run proceeds independently until convergence to a locally optimal solution. The best solution among all runs is then selected as the final result.

\vspace{2pt}
\noindent
\textbf{Optimization Objective.}
The optimization objective that we use minimizes the total workload cost as defined earlier (Section~\ref{subsec:cost_estimation}):
\vspace{-5pt}
\[
\min_{S \in \mathcal{S}_{\text{valid}}} 
\left(
\sum_{T_i \in \mathcal{T}} w_i \cdot C(T_i, S)
+
\sum_{j} I_j \cdot C_{\text{ins}}(j,S),
\right).
\]
\vspace{-15pt}

\vspace{5pt}
\section{Implementation}
\label{sec:implementation}

Figure \ref{fig:erbiumdb} presents the high-level architecture of ErbiumDB. The system currently runs on top of a relational backend (PostgreSQL), but its architecture is
extensible and can support additional backends. 
The user-provided E/R schema is processed by the Schema DDL Parser which constructs an internal E/R graph representation in which each entity set, relationship set, and attribute is modeled as a node enriched with structural metadata. The resulting graph forms the basis for generating all metadata required for conceptual-to-relational schema compilation.  The Query Workload Analyser extracts query templates and
their corresponding frequencies from the input workload, using which the Mapping optimizer searches for the minimum-cost relational schema, which is then materialized in the underlying backend.
To preserve logical data independence, the E/R graph and all associated mapping metadata are serialized and stored as a system table in the database. This enables reconstruction of the conceptual abstraction and the parameters of the chosen physical mapping through deserialization. The Query Compiler transparently rewrites 
CRUD SQL queries defined over the conceptual schema into equivalent SQL queries over the optimized relational schema. Currently only a limited set of queries are supported. From the user’s perspective, the E/R model serves as the logical schema, while the underlying relational representation remains hidden, thereby providing full logical data independence.

\subsection{Query Example}
\label{subsec:query_example}

We express E/R queries in SQL-like syntax, with each entity
set and relationship set exposed as a separate logical
relation. References to entity sets are polymorphic.
For example, \texttt{Customer} includes its direct
instances and instances of all its subtypes.
Consider a workload consisting of the following E/R query which returns the ID, name, and distinct software count of each
customer who has downloaded at least five software items
satisfying the license predicate.
\begin{lstlisting}[
  language=SQL,
  basicstyle=\small\ttfamily,
  breaklines=true,
  breakatwhitespace=true,
  columns=fullflexible,
  keepspaces=true,
  showstringspaces=false,
  linewidth=\columnwidth
]
SELECT e2.user_id, e2.name, 
       COUNT(DISTINCT e1.p_id) AS related_count
FROM software_downloads AS r
JOIN software AS e1
  ON r.p_id = e1.p_id
JOIN customer AS e2
  ON r.user_id = e2.user_id
WHERE e1.license_type = 'open_source'
GROUP BY e2.user_id, e2.name
HAVING COUNT(DISTINCT e1.p_id) >= 5;
\end{lstlisting}
The workload abstraction assigns a frequency of one to
the \texttt{SELECT *} templates for
\texttt{Software\_downloads}, \texttt{Software}, and
\texttt{Customer}. All other select frequencies and all
insert counts are zero. Greedy uses this profile to select
a mapping. The compiler then translates the original query
for that mapping, and we measure its execution time in
PostgreSQL.

Based on the selected the relational schema, E/R query is rewritten to corresponding relational query. 
In the extended schema that we considered for motivating experiment (Section~\ref{sec:motivating-experiment}), \texttt{Customer} has two disjoint
subtypes, \texttt{PrimeCustomer} and
\texttt{BusinessCustomer}. 
Under modified concrete-table
inheritance mapping, the three types are stored separately in
\texttt{relation\_1}, \texttt{relation\_2}, and
\texttt{relation\_3}, respectively. The download
relationship and software entities are stored in
\texttt{relation\_4} and \texttt{relation\_5}. The
three customer tables are combined to resolve
the polymorphic customer endpoint.
\begin{lstlisting}[
  language=SQL,
  basicstyle=\small\ttfamily,
  breaklines=true,
  breakatwhitespace=true,
  columns=fullflexible,
  keepspaces=true,
  showstringspaces=false,
  linewidth=\columnwidth
]
WITH customers(user_id, name) AS (
    SELECT customer_id, name FROM relation_1
    UNION ALL
    SELECT primecustomer_id, name FROM relation_2
    UNION ALL
    SELECT businesscustomer_id, name FROM relation_3
)
SELECT c.user_id, c.name,
       COUNT(DISTINCT s.software_id) AS related_count
FROM relation_4 AS d
JOIN relation_5 AS s
  ON d.software_id = s.software_id
JOIN customers AS c
  ON d.customer_id = c.user_id
WHERE s.license_type = 'open_source'
GROUP BY c.user_id, c.name
HAVING COUNT(DISTINCT s.software_id) >= 5;
\end{lstlisting}

Under single-table inheritance mapping, the Product and User
hierarchies are stored in \texttt{relation\_2} and
\texttt{relation\_3}, respectively, and the download
relationship is stored in \texttt{relation\_1}. The
rewritten query uses type predicates on \texttt{role} to select
software rows from the Product table and customer rows,
including both subtypes, from the User table.
\begin{lstlisting}[
  language=SQL,
  deletekeywords={ROLE},
  basicstyle=\small\ttfamily,
  breaklines=true,
  breakatwhitespace=true,
  columns=fullflexible,
  keepspaces=true,
  showstringspaces=false,
  linewidth=\columnwidth
]
SELECT u.user_id, u.name,
       COUNT(DISTINCT p.product_id) AS related_count
FROM relation_1 AS d
JOIN relation_2 AS p
  ON d.software_id = p.product_id
JOIN relation_3 AS u
  ON d.customer_id = u.user_id
WHERE p.role = 'software'
  AND p.license_type = 'open_source'
  AND u.role IN ('customer', 'primecustomer',
                 'businesscustomer')
GROUP BY u.user_id, u.name
HAVING COUNT(DISTINCT p.product_id) >= 5;
\end{lstlisting}

\section{Experimental Evaluation}
\label{sec:experiments}

We first describe the data generation process used to construct the experimental datasets and workloads, as suitable real-world workloads that capture the diversity of conceptual-to-relational mappings are not readily available. We conduct experiments 
on both synthetically generated and real-world-inspired E/R schemas. All experiments are performed on a machine with 20 CPU cores and 32~GB of RAM.
We create only primary-key indexes on the generated relational tables.

\subsection{Data Generation}
\label{subsec:data_gen}

We use the same E/R schema as used in the motivating experiment in Section~\ref{sec:motivating-experiment}.
We define a node count for each entity set in the E/R schema. For entity sets that participate in an inheritance hierarchy, the node count represents the number of tuples that belong exclusively to that entity set. Node counts for leaf entity sets in a hierarchy are generated according to a truncated Zipf distribution with parameter $\alpha$. 
Leaf entity sets are first assigned a random ordering, which determines their ranks. If a leaf entity set is assigned rank $k$, the probability associated with that rank is given by
\[
p(k) = \frac{1/k^{\alpha}}{\sum_{j=1}^{N} 1/j^{\alpha}},
\]
where $N$ denotes the number of leaf entity sets in the hierarchy. The leaf entity set with the lowest probability (rank $k=N$) is assigned a node count of $10^5$. Node counts for the remaining leaf entity sets are scaled proportionally to their Zipf probabilities, yielding
\[
N_k = 10^5 \cdot \frac{p(k)}{p(N)}, \quad k = 1, \dots, N,
\]
where $N_k$ denotes the node count of the entity set assigned rank $k$.
 For each non-leaf (parent) entity set, the node count is assigned as a skewed value computed as the sum of the node counts of its descendant entity sets multiplied by a randomly chosen factor in the range [0.005,2], with the distribution of the factor heavily skewed toward 0.005. For non-hierarchical entity sets, the node count is randomly selected from the range $(10000, 150000)$ with higher probability towards max value. For hierarchical entity sets, the cardinality is computed as the sum of its own node count and the node counts of all its descendant entity sets. For non-hierarchical entity sets, the cardinality is equal to the node count. 

For each relationship set in the E/R schema, we define a participation factor. For many-to-one relationship sets, this factor denotes the percentage of tuples from the many-side entity set that participate in the relationship set. For one-to-one and many-to-many relationship sets, it denotes the percentage of tuples from the first entity (in the order specified by the input ER schema) that participate in the relationship. For M:N relationships, we additionally define an entity1-to-entity2 factor, which specifies the expected number of matches in the second entity per participating tuple in the first entity. Specifically, for each participating tuple in entity1, the number of matching tuples in entity2 is drawn uniformly at random from the interval $(1,f]$, where $f$ is the entity1-to-entity2 factor. The resulting relationship cardinality is derived from both factors for many-to-many relationship sets, and from the participation factor alone for one-to-one and many-to-one relationship sets. For multi-valued attributes, we define a parameter that specifies the expected number of attribute values generated per participating entity or relationship tuple. For each participating tuple, the number of attribute values is randomly chosen from the range $[1, 2 \times \text{parameter}]$.

\subsection{Is the Greedy Optimizer Workload-Aware?}

\begin{table}[t]
\centering
\caption{Measured execution times and Greedy speedups.
For each workload, the cost model selects the configuration
with the lowest estimated cost among the six fixed
configurations. Letters in parentheses identify
the configurations in Table~\ref{tab:fixed-configurations}.}
\label{tab:greedy-comparison}
\footnotesize
\setlength{\tabcolsep}{2.5pt}
\begin{tabular}{@{}lrrrrr@{}}
\toprule
& \multicolumn{3}{c}{Measured execution time (s)}
& \multicolumn{2}{c}{Greedy speedup over} \\
\cmidrule(lr){2-4}\cmidrule(l){5-6}
Workload &
\shortstack{Fastest\\fixed} &
\shortstack{Model-selected\\(among 6 fixed)} &
Greedy &
\shortstack{Fastest\\fixed} &
\shortstack{Slowest\\fixed} \\
\midrule
W1  & 225.7 (E) & 244.6 (F) & 220.9 & 1.02 &  2.50 \\
W2  &  13.7 (A) &  14.8 (B) &  11.7 & 1.17 & 20.56 \\
W3  & 240.5 (F) & 240.5 (F) & 217.9 & 1.10 &  2.48 \\
W4  & 264.3 (E) & 301.4 (F) & 260.3 & 1.02 &  2.70 \\
W5  &  12.3 (A) &  13.2 (B) &  13.0 & 0.95 & 21.65 \\
W6  & 213.8 (D) & 280.3 (F) & 214.0 & 1.00 &  2.28 \\
W7  & 147.6 (E) & 159.7 (F) & 114.9 & 1.28 &  2.09 \\
W8  &  12.6 (A) &  18.2 (B) &  14.6 & 0.86 & 15.86 \\
W9  &  16.1 (B) &  16.1 (B) &  15.6 & 1.03 & 16.31 \\
W10 & 199.8 (E) & 203.9 (F) & 195.3 & 1.02 &  2.63 \\
\bottomrule
\end{tabular}
\vspace{-10pt}
\end{table}

\vspace{4pt}\noindent
\textbf{Experiment 1.}
We evaluate the greedy optimizer using the same ten complex E/R query
workloads and experimental setup as in the motivating
experiment (Section~\ref{sec:motivating-experiment}).
For each workload, we score the six fixed configurations
in Table~\ref{tab:fixed-configurations} using our cost model, and select the one
with the lowest estimated cost. We then use the same
cost model with the greedy optimizer, 
rewrite the workload's
queries against the schema selected by greedy search, execute
them in PostgreSQL, and measure their total execution
time. For the 6 fixed configurations, we reuse the measured
runtimes from the motivating experiment. For each
workload, we compute Greedy's speedup over the fastest
and slowest fixed configurations by dividing their
respective runtimes by Greedy's runtime.

Table~\ref{tab:greedy-comparison} shows that the cost
model selects the fastest fixed configuration or comes
with 15\% of its runtime for 8 out of 10 workloads, and
is within 50\% for the remaining two workloads. 
This illustrates that our cost model, despite simplistic assumptions, is 
quite effective at identifying good mappings.
Greedy outperforms the model-selected fixed configuration
on all ten workloads and the fastest measured fixed
configuration on seven, achieving up to $1.28\times$ speedup over
the fastest fixed configuration on W7. On the remaining
workloads, Greedy closely matches the fastest fixed
configuration on W6 and incurs approximately 5\% and
15\% additional execution time on W5 and W8, respectively. Relative to the slowest fixed configuration, Greedy
achieves speedups ranging from $2.09\times$ to
$21.65\times$. These results indicate that workload-aware
mapping selection produces designs that are generally
competitive with the best evaluated fixed configuration
while avoiding the substantial execution-time penalties
of poorly suited mappings.

\vspace{4pt}\noindent
\textbf{Experiment 2:}
For each workload, we randomly select a subset
$\mathcal{C}' \subseteq \mathcal{C}\setminus\mathcal{M}$
containing approximately one-sixth of the entity and
relationship sets. For each component
$c \in \mathcal{C}\setminus\mathcal{M}$, we assign
\[
w_c = I_c =
\begin{cases}
1, & c \in \mathcal{C}',\\
0, & \text{otherwise}.
\end{cases}
\]
We generate 12 workloads using this procedure, each
focusing on a randomly selected subset of schema components.
\begin{figure}[t]
  \centering
  \hspace*{-1.5mm}%
  \includegraphics[scale=0.14,
    trim=3cm 3cm 1cm 1.5cm,
    clip]{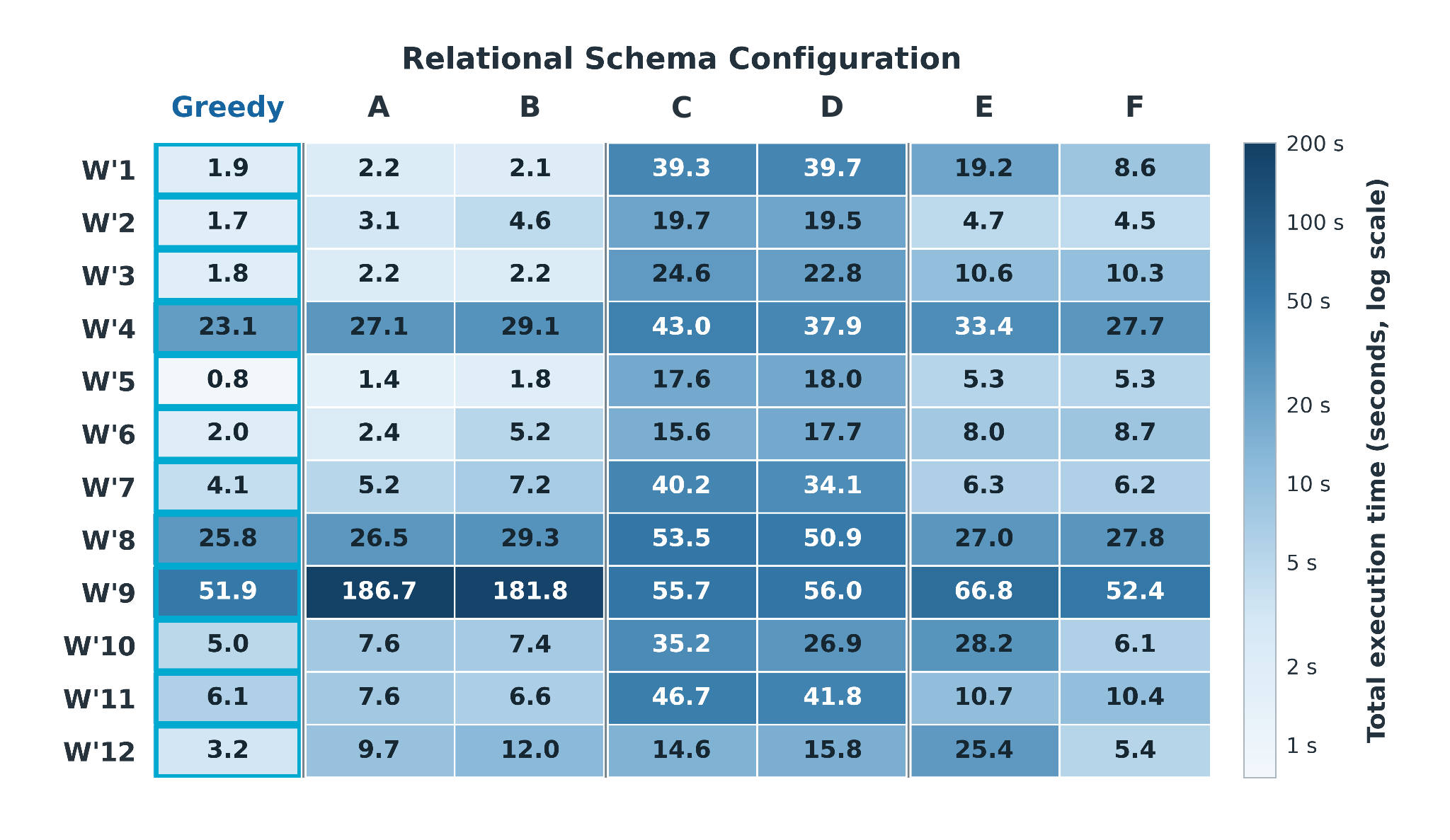}
  \caption{Total PostgreSQL exec time (s) per workload (Expt 2).}
  \label{fig:heatmap-ecommerce-12-workloads}
  \vspace{0pt}
\end{figure}
Figure~\ref{fig:heatmap-ecommerce-12-workloads} compares
the total PostgreSQL execution time under Greedy and
the six fixed configurations defined in
Table~\ref{tab:fixed-configurations}.
Greedy achieves the lowest execution time on all
12 workloads. Relative to Greedy, the maximum slowdowns
of the modified concrete-table variants (Configs.~A--B),
class-table variants (Configs.~C--D), and single-table
variants (Configs.~E--F) reach approximately
$3.7\times$, $21.2\times$, and $10.0\times$, respectively.
These results demonstrate the benefits of selecting
mappings at the component level according to workload
demand, particularly when the workload focuses on a
small subset of the schema.

\subsection{Is the cost model predictive?}

\begin{figure}[h]
  \centering
  \includegraphics[scale=0.46,
  trim=0cm 0cm 0cm 0cm,
  clip]{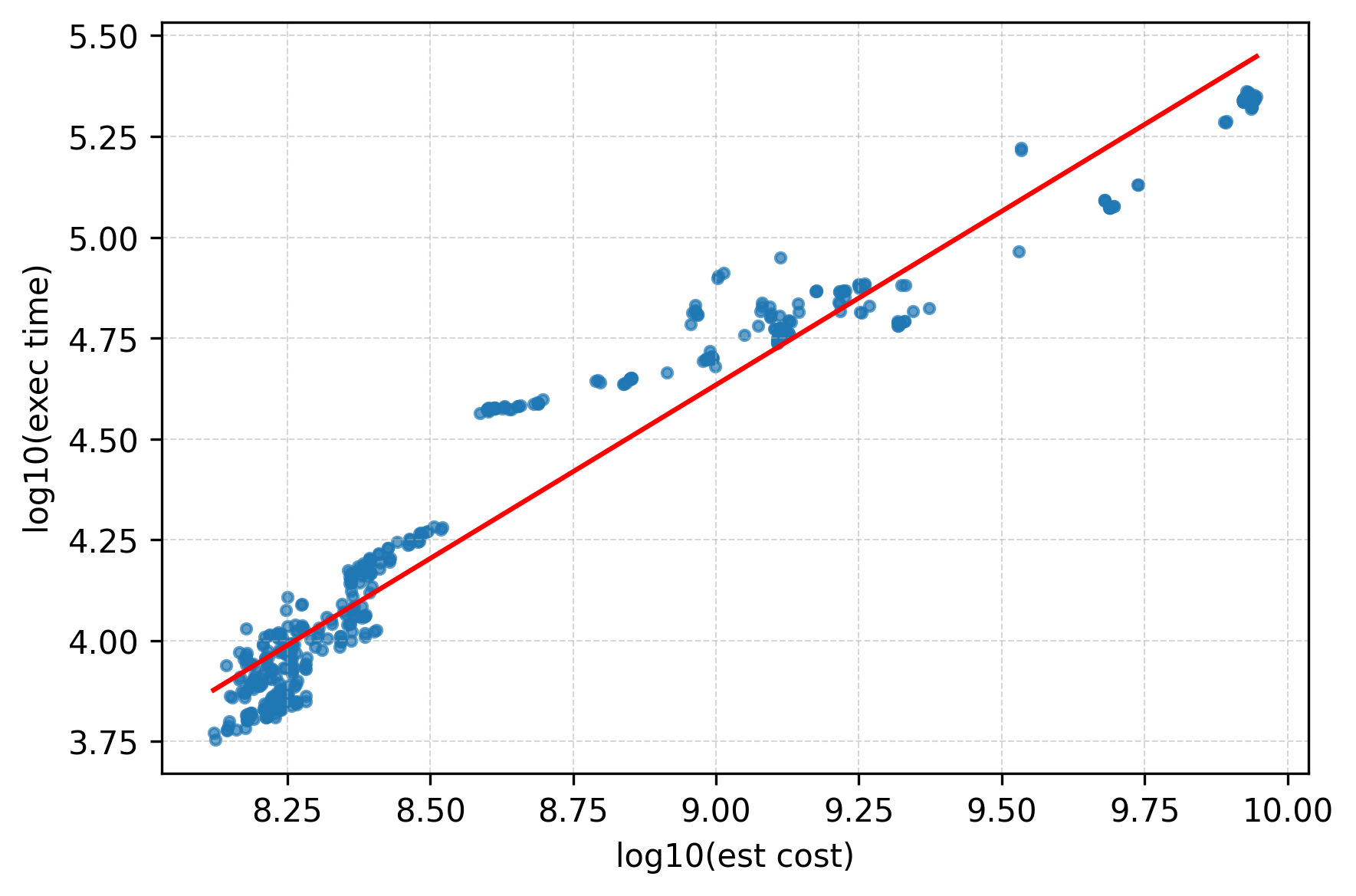}
  \caption{Estimated Cost vs. PostgreSQL Execution Time (Log–Log Scale) for 500 randomly chosen relational schema configurations.}
  \vspace{0em}
  \label{fig:cost-model}
\end{figure}

We generate workload patterns by varying the weight vector $\mathbf{w}$ associated with the query templates $\mathcal{T}$ and the insert workload counts $\mathcal{I}_i$ for each component $c_i \in \mathcal{C}$ in the schema. For select queries, each query is executed five times, and the median execution time is reported as the execution time of the query.
We evaluate the accuracy of the cost model by comparing the estimated cost of a schema $S$ with its measured total execution time across 500 randomly sampled relational schema configurations from the mapping space. For each configuration, we plot the estimated cost against the observed execution time. As shown in Figure~\ref{fig:cost-model}, the estimated cost exhibits a strong positive correlation with the observed execution time on a log-log scale. The data points align closely with the fitted linear trend, indicating that the cost model effectively captures the relative differences in execution time across schema configurations. While some dispersion is observed due to simplifications in the cost model, the overall trend demonstrates that the model is sufficiently predictive to guide the optimization process.

\section{Related Work}

Research on higher-level data abstractions spans E/R models,
property graphs, XML and RDF, and object-oriented schemas.
Prior work has studied entity-centric abstractions and the
connections between E/R schemas and property
graphs~\cite{chen1976entity, E-R-graphs}. A substantial body of
work has formalized graph query languages and studied their
execution over relational
backends~\cite{angles2018foundations,pg-keys,jindal2014graphiq,
opencypher2017,gsparql2012,sqlgraph}. Related work has explored
relational representations for XML and RDF~\cite{rdf-3x,
xml-querying-relational} and translations from E/R to
XML~\cite{conceptual-xml,conceptual-xml-graph}. LegoDB is a
close precedent for logical design tuning over a relational
backend. It uses workload cost estimates to explore alternative
XML-to-relational mappings~\cite{legodb}.

Object-relational mapping frameworks such as
Hibernate~\cite{hibernate} expose object models over relational
storage~\cite{impedance}. Prior work has studied flexible
inheritance mappings and the translation of CRUD
operations~\cite{morm}. The ADO.NET Entity Framework similarly
exposes a conceptual schema through declarative
mappings~\cite{adonet}. Related work studies the compilation
of these mappings into bidirectional transformations and
query containment in Entity
SQL~\cite{mapping_compilation,query_containment}. 
These approaches typically use fixed mappings or select them using heuristics.
In contrast, Bagheri et al.~\cite{bagheri2017orm} generate alternative ORM designs from formal specifications and benchmark them by executing synthesized workloads. Later learning-based approaches use performance predictions
to guide exploration of ORM designs~\cite{DTS,y-map,y-map-2}.
TriORM uses predicted query and write costs to recommend ORM
mappings for a given workload~\cite{triorm}.

Traditional relational design generally derives a normalized logical schema from a conceptual model and uses workload information primarily to select physical structures.
Workload-aware physical design has been studied extensively
for index selection~\cite{index-selection-surajit,
index-selection-bruno,das2019automatically,
index-selection-schlosser,Idreos2007DatabaseC,autoadmin},
partitioning~\cite{partitioning-hammer,partitioning-rao,
partitioning-jigsaw,partitioning-bruno,partitioning-grid},
and materialized views~\cite{views-gupta,views-ullman,
views-yang,views-yuan,views-chuan}. Mozaffari et al. survey
this literature~\cite{mozaffari2024selftuning}.
By contrast, workload-driven logical design has been explored for
NoSQL systems. NoSE uses a conceptual model and query
workload to select a Cassandra schema~\cite{NOSE}. Other
approaches derive denormalized representations from
relational schemas~\cite{vajk2013denormalizing}, apply
query-driven mapping rules~\cite{chebotko2015cassandra}, or
generate and rank alternative column-store schemas for
known queries~\cite{bermbach2015informed}. These approaches target NoSQL-specific representations, whereas our framework performs workload-aware selection among semantically equivalent E/R-to-relational mappings for a relational backend.

\section{Conclusion}

Although substantial prior work has focused on optimizing physical design, the logical design decisions that govern how entities, relationships, and inheritance hierarchies are represented in relational schemas have received comparatively less attention, despite their significant impact on performance. As a result, these performance-critical decisions are often left to users. We present a systematic approach that elevates the user-facing layer of the database to a widely used conceptual abstraction, namely the E/R model, while automatically deriving relational schemas in a workload-aware manner. Our framework treats conceptual-to-relational mapping decisions as first-class design variables and optimizes them based on workload characteristics. Our experimental results show that these schema design choices directly affect the efficiency of query processing. In particular, fixed or heuristic mapping strategies fail to adapt to diverse workloads and application-specific schemas, whereas our workload-aware approach consistently achieves significant performance improvement across a wide range of scenarios.

There are several directions for future work that we are planning to pursue. 
We plan to build on our preliminary work on investigating mechanisms for enforcing constraints specified at the E/R abstraction within the relational backend~\cite{cidr-paper}. 
It would also be interesting to study joint optimization of logical and physical design choices, such as combining mapping selection with index and materialized-view selection.  
In particular, we aim to understand how these physical structures interact with mapping choices and whether joint optimization can narrow performance gaps among alternative relational realizations while improving the robustness of logical designs to workload variation.
Finally, we plan to explore dynamic re-optimization as workloads evolve over time, and to extend the framework to other high-level abstractions beyond E/R models.

\section{AI-Generated Content Acknowledgment}

The authors used OpenAI ChatGPT to assist with 
debugging the research prototype, setting up experiments, and
refining and editing portions of the manuscript text for clarity and presentation.

\bibliographystyle{unsrt}
\bibliography{sample-base}

\end{document}